\documentstyle[prl,aps,multicol,epsf]{revtex}

\begin{document}

\newcommand {\be}[1] {\begin{equation}\label{#1}}
\newcommand {\ee} {\end{equation}}
\newcommand {\bea}[1] {\begin{eqnarray}\label{#1}}
\newcommand {\eea} {\end{eqnarray}}
\def \OMIT #1{}
\def \rem #1 {{\it #1}}
\newcommand {\Eq}[1] {Eq.~(\ref{#1})}
\newcommand {\Fig}[1] {Fig.~\ref{#1}}

\title {Critical geometry of two-dimensional passive scalar turbulence}

\author{Jan\'{e}  Kondev}
\address{Physics Department, Brandeis University, Waltham, MA 02454}

\author{Greg Huber}
\address{Department of Physics, University of Massachusetts - Boston, 
100 Morrissey Boulevard, Boston, MA 02125}
             
\date{\today}

\maketitle

\widetext

\begin{abstract}
Passive scalars advected by a magnetically driven two-dimensional 
turbulent flow are analyzed using methods of statistical topography. 
The passive tracer concentration is interpreted as the height of a 
random surface and the scaling properties of its contour loops are analyzed. 
Various exponents that describe the loop ensemble are measured and 
compared to a scaling theory. This leads to a geometrical criterion for the  
intermittency of scalar fluctuations. 
\end{abstract}

\pacs{PACS numbers:  }

\begin{multicols}{2}

New experimental techniques which provide complete real-space 
images of fluctuating fields are being developed throughout 
the physical sciences \cite{PC,gollub_old,putter}. 
They provide data on spatial structures 
inaccessible to methods based on system-wide averages, or on measurements 
that probe the system at only a few points. 
Consequently, new measures designed to usefully exploit this wealth of 
information 
are of great practical importance. Developing such measures not only provides
the experimentalist with new tools with which to analyze and classify 
correlations  in image data, but may also 
inspire theories that can predict the correlation behavior.

Scalar turbulence,  which deals with dispersion of a passive 
substance (pollutants, temperature, etc.) by a turbulent 
flow,  provides an intriguing example of this paradigm. When describing the 
fluctuations of the tracer concentration,  
recently the focus has shifted from the so-called structure functions, 
which characterize the probability distribution function  for the 
concentration difference between two points,  
to multi-point correlations which better take into account spatial information. 
This has lead to significant advances in turbulence research. 
In particular it has 
allowed for the identification of the source of {\em intermittency} of the 
scalar fluctuations, i.e.,   
the breakdown of simple scaling as dictated by dimensional analysis \cite{siggia}. 
This shift of perspective is nicely exemplified by a recent study of the Lagrangian 
evolution of three tracer particles \cite{vergasola}. It was shown 
that intermittency in the structure functions can be related to the 
evolution of the {\em shape} of the triangle defined by the three particles, 
while the overall size of the triangle evolved with a simple scaling law. 

Here  we develop a geometrical description  
of  scalar fields dispersed by two-dimensional turbulence,  
based on their level sets.   Unlike earlier studies of the fractal properties of
the level set \cite{PC,gollub_old,tabeling_old}, our work
focuses on distribution functions of level-set loops, in analogy with
cluster distributions at a critical point.  Similar ideas were exploited recently by 
Catrakis and Dimotakis \cite{greeks} who introduced the  joint distribution function of
shape complexity and size to describe level-set 
islands and lakes obtained from images of jet turbulence. 
We define new  measures  for the random geometry defined by 
the scalar field and propose scaling 
relations between various exponents that characterize these measures.  
Checking for the validity of these scaling relations reveals interesting 
properties of the fluctuations of the scalar field, in particular it provides a new, 
geometrical way of identifying intermittency in experimental scalar turbulence data.

\paragraph{{\bf Scaling theory of contour loops}} 

The scaling theory of contour loops of {\em self-affine} rough fields (defined below) 
was developed in Refs.~\cite{isich} and~\cite{JK00}. Here we extend this theory 
to analyze the scaling properties of scalar 
fields advected by turbulent flows, which are typically intermittent, 
i.e., {\em not} self-affine. 
For this purpose we distinguish between scaling relations which are a 
consequence of the scale invariance of the loop ensemble only, and those 
which require the self-affine property for the concentration field as well. 
This allows us to test these two distinct hypothesis separately  
against experimental data. 

For a random two-dimensional field $c({\bf x})$,  
we define the contour loop ensemble as a union of level sets for different 
values of $c$.  Each level set consists of contour loops which come in  
many shapes and sizes. The contour loop ensemble 
is characterized by the loop correlation function $G({\bf r})$,  
which measures the 
probability that two points separated by ${\bf r}$ 
are on the {\em same} contour loop, and the joint distribution function of 
loop lengths and radii $n(s,R)$. Here $n(s,R) \: ds dR$ is the number of loops, 
per unit area, which pass through a fixed point (say the origin), and 
whose  length $s\in (s, s+ds)$, and radius $R\in (R, R+ dR)$. The radius of a 
loop is defined by the side of the smallest square that covers it. 

Assuming that the contour loop ensemble is {\em self-similar}, we can define 
{\em geometrical exponents} which express the scaling 
properties of the two measures, $G({\bf r})$ and $n(s,R)$. The loop correlation 
function, 
\be{loopcorr}
  G(r) \sim \frac{1}{r^{2 x_l}} \; ,
\ee
defines the loop correlation exponent $x_l$, while the joint distribution 
function has a scaling form, 
\be{nsal}
n(s,R)
\sim s^{1-\tau-1/D} \: f_{n}(s/R^{D})  \ ,
\ee 
and it  defines the exponents $D$ and $\tau$. 
Both relations are expected to 
hold for large loops with radii much bigger than the lattice spacing 
(for image data, lattice spacing = pixel size).  
For a self-similar contour loop ensemble $D$ is also the 
fractal dimension of large loops, not too be confused with the dimension of the 
whole level set studied previously \cite{PC}.

These three geometrical exponents satisfy a scaling relation, 
\be{scalerel1} 
D (3- \tau) = 2 - 2 x_l
\ee
which is derived from a sum rule \cite{JK00}. The sum rule equates the two ways of  
calculating the mean  loop length in a finite region, one 
using $G(r)$ and the other based on  $n(s,R)$. This scaling relation is a 
useful check on the validity of the assumption of scale invariance of the loop 
ensemble and in particular Eqs.~\ref{loopcorr}~and~\ref{nsal}.

For a random field which is rough ($0\le\alpha\le1$) 
{\em and} self affine we can derive a second relation: 
\be{scalerel2}
D (\tau - 1) = 2 -\alpha
\ee
where $\alpha$ is the {\em affine exponent}. It is defined by the relation   
$c({\bf x}) \equiv b^{-\alpha} c(b {\bf x})$,   
where ``$\equiv$'' means that the original field and its rescaled version  
are statistically equivalent; $b>1$
is an arbitrary scale parameter. \Eq{scalerel2}  
is a kind of hyperscaling relation, and it follows from the fact that the 
number of large loops at a particular scale of observation grows with the 
scale to power $\alpha$ \cite{JK00}. (The usual hyperscaling relation, $D(\tau-1)=2$, 
say in 2D critical percolation, follows from the assumption that at every scale
the number of large clusters  is of order one, i.e., $\alpha=0$.) 

Furthermore, for self-affine rough fields the three geometrical
exponents, $\tau$, $D$, and $x_l$,
depend on the value of $\alpha$ {\em only} \cite{isich}. Namely, based on  
exact results derived in the two limiting
cases $\alpha=0$ and $\alpha=1$,  $x_l=1/2$ independent of $\alpha$ was
conjectured \cite{JK00}; this has been checked in
numerical simulations \cite{JK00,zeng}. 
Once this value of $x_l$ is accepted, formulas 
for the other two geometrical exponents, 
\be{D-alpha}
    D = \frac{3 - \alpha}{2} \ ,   \; \; \; \;
    \tau - 1 = \frac{4 - 2\alpha}{3 - \alpha} \  , 
\ee
follow from \Eq{scalerel1} and \Eq{scalerel2}. 

We can  analyze the level sets of random two-dimensional fields based on the 
above scaling theory. 
The procedure consists of tracing out contour loops of
$c({\bf x})$. Then for each loop its radius and length are
recorded,  as well as its  contribution to the loop correlation function.
The three geometrical exponents are measured by power-law fitting of
$G(r)$ (for $x_l$), the average loop length as a function of radius
(for $D$), and the number of loops whose length is greater than $s$ (for $\tau-2$).
The last two measures are derived from $n(s,R)$ by 
integration:
\bea{int} 
\left<s\right>(R) & \equiv &  \frac{\int_0^\infty\! ds \: n(s,R) s}{ 
 \int_0^\infty\!ds \:  n(s,R)} \ \sim R^D \nonumber  \\ 
 N_>(s) & = &  \int_s^\infty\!ds' \int_0 ^ \infty\!dR \: n(s',R) \ \sim s^{-(\tau-2)} 
\ .
\eea
Given that we can reliably extract the three exponents (i.e. the data 
show at least a decade of scaling) we can test our assumptions about the 
loop ensemble by checking the validity of the various scaling relations. 
First, if the loop ensemble is self similar, \Eq{scalerel1} should hold. 
Furthermore, if it is self-affine, we expect both relations in \Eq{D-alpha} to 
hold. 
We check this by extracting $\alpha$ from $D$ and $\tau$ separately 
and comparing them. For an intermittent field we do not expect the two 
values of $\alpha$ to coincide.

\paragraph{{\bf Experimental data}}

The contour loop analysis is performed on images of a fluorescent dye (passive
scalar) advected by magnetically forced two-dimensional turbulent flows 
\cite{tab_batch,tab_enst}.
The flow is generated by running a current through a thin slab of salty water in 
the presence of a perpendicular magnetic field; the Lorentz force 
on the charged ions provides the stirring. 
The magnetic field is generated by an array of permanent 
magnets placed in proximity to the flow. 

With a suitable arrangement of magnets and choice of driving current a 
turbulent velocity field is set up either in the regime of a 
direct enstrophy cascade 
\cite{tab_batch} 
or an inverse energy cascade \cite{tab_enst}. In both 
cases the advection of fluorescent dye was monitored using a CCD 
camera. The data we analyze below consists of two sets, one for each of the 
two flow regimes, of 10 images of the tracer field. These images were 
selected from the time interval during 
which stationarity of the tracer fluctuations was achieved. In this time interval the  
energy dissipation was observed to be roughly time-independent \cite{tab_batch}.

\paragraph{{\bf Results of the loop analysis}} 

The contour loop analysis for the two types of forcing was performed 
by measuring the loop correlation function, the average loop length  
as a function of the loop radius, and the distribution of loop lengths from the images 
provided by P.~Tabeling's  group. 
Each image consists of a $512\times 512$ array of integers which represent the 
gray scale values of the intensity of the fluorescent dye, which is proportional to its  
concentration $c({\bf x})$. The unit of length we use throughout is the lattice spacing
which physically corresponds to the size of one pixel. 
For each 
image $10^4$ points on the dual lattice were chosen at random  
and through each point a contour loop 
was constructed following the algorithm described in \cite{JK00}. 
For each loop we measured the length and radius, and 
its contribution  to the loop correlation function. 

To check the stationarity condition for the data from the two flow regimes 
we performed the loop 
analysis on each image separately. We found no discernible variation in time 
for the three measures $G({\bf r})$, $\left<s\right>(R)$, and $N_>(s)$. Therefore we
combined the loop data from all 10 images to improve  our statistics.

\end{multicols}
 
\begin{minipage}[t]{5cm}
\begin{figure} \label{plots}
\epsfxsize=5cm \epsfbox{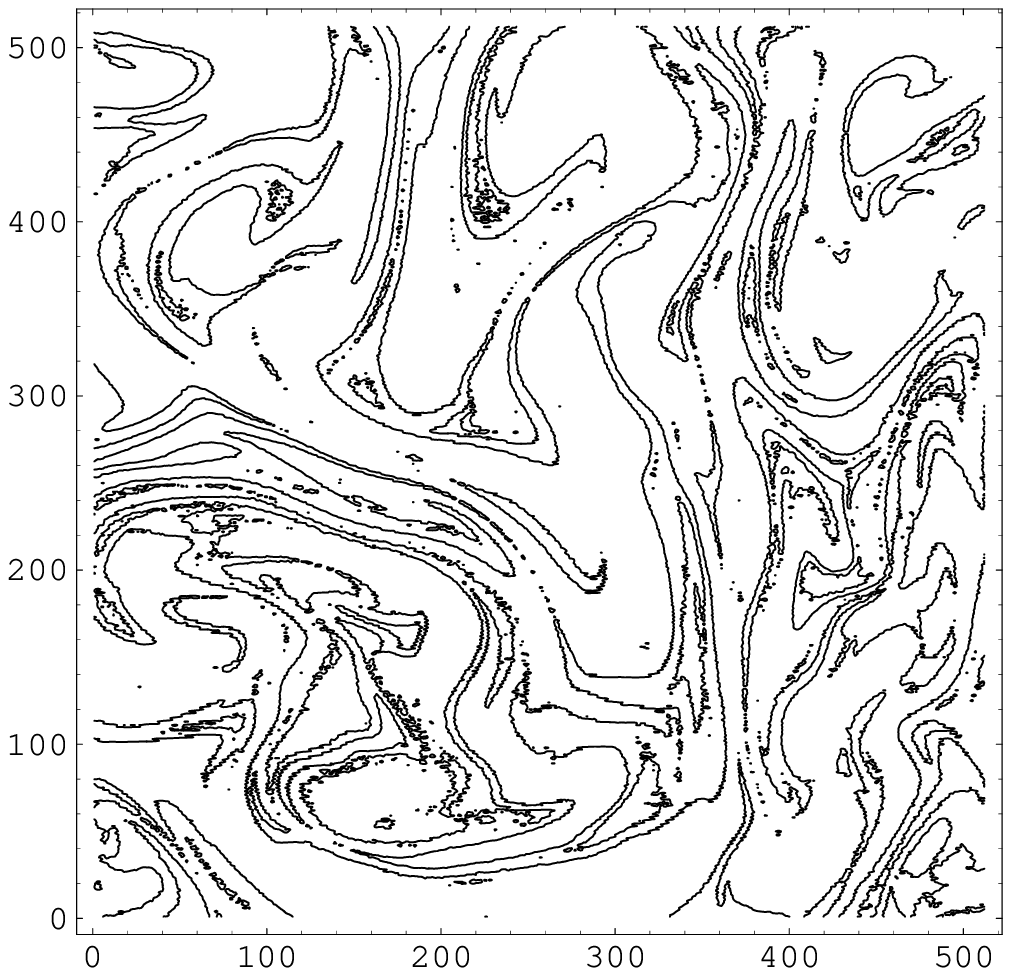}
\end{figure}
\end{minipage}
\hfill
\begin{minipage}[t]{6cm}
\begin{figure}
\epsfxsize=6cm \epsfbox{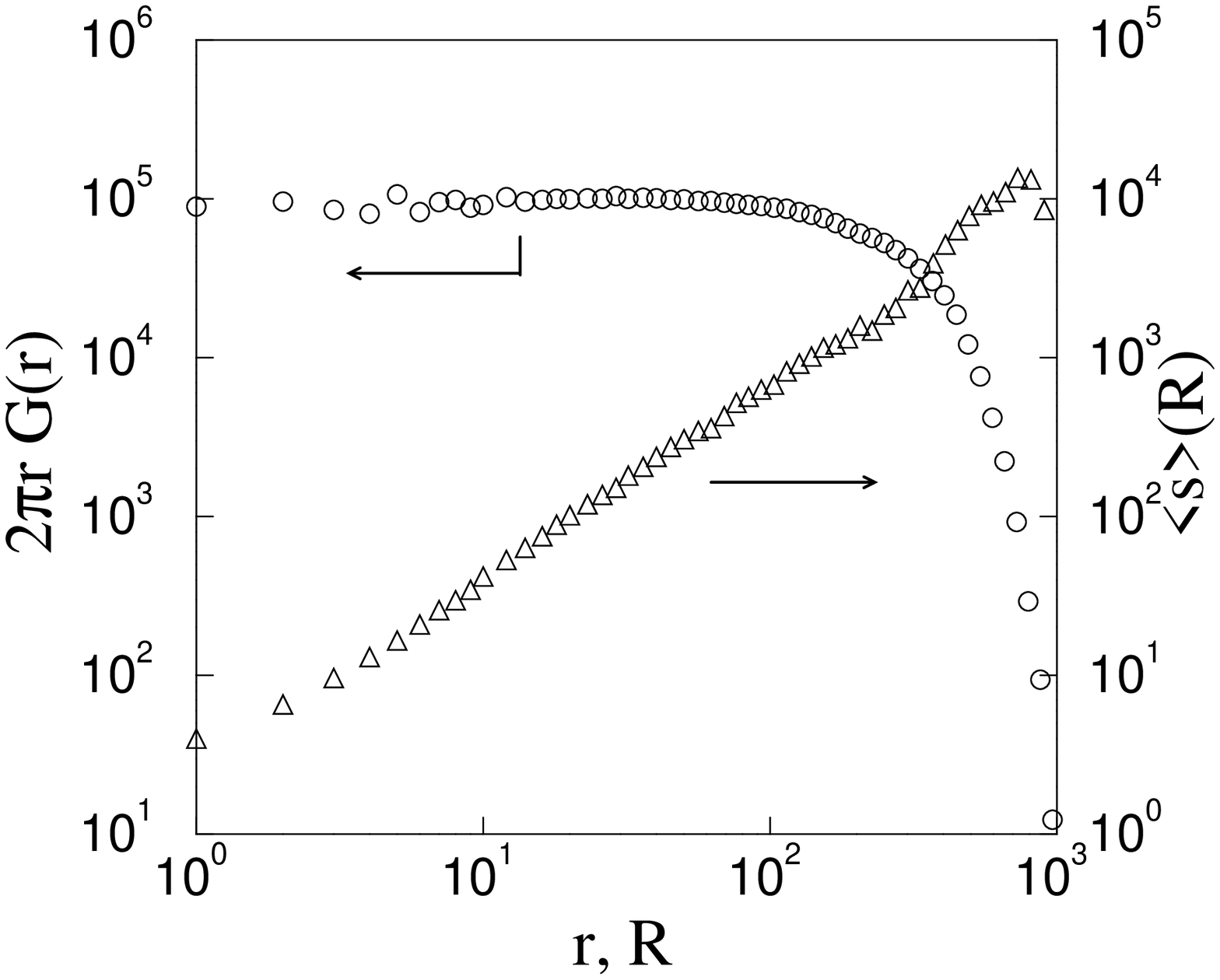}
\end{figure}
\end{minipage}
\hfill
\begin{minipage}[t]{6cm}
\begin{figure}
\epsfxsize=6cm \epsfbox{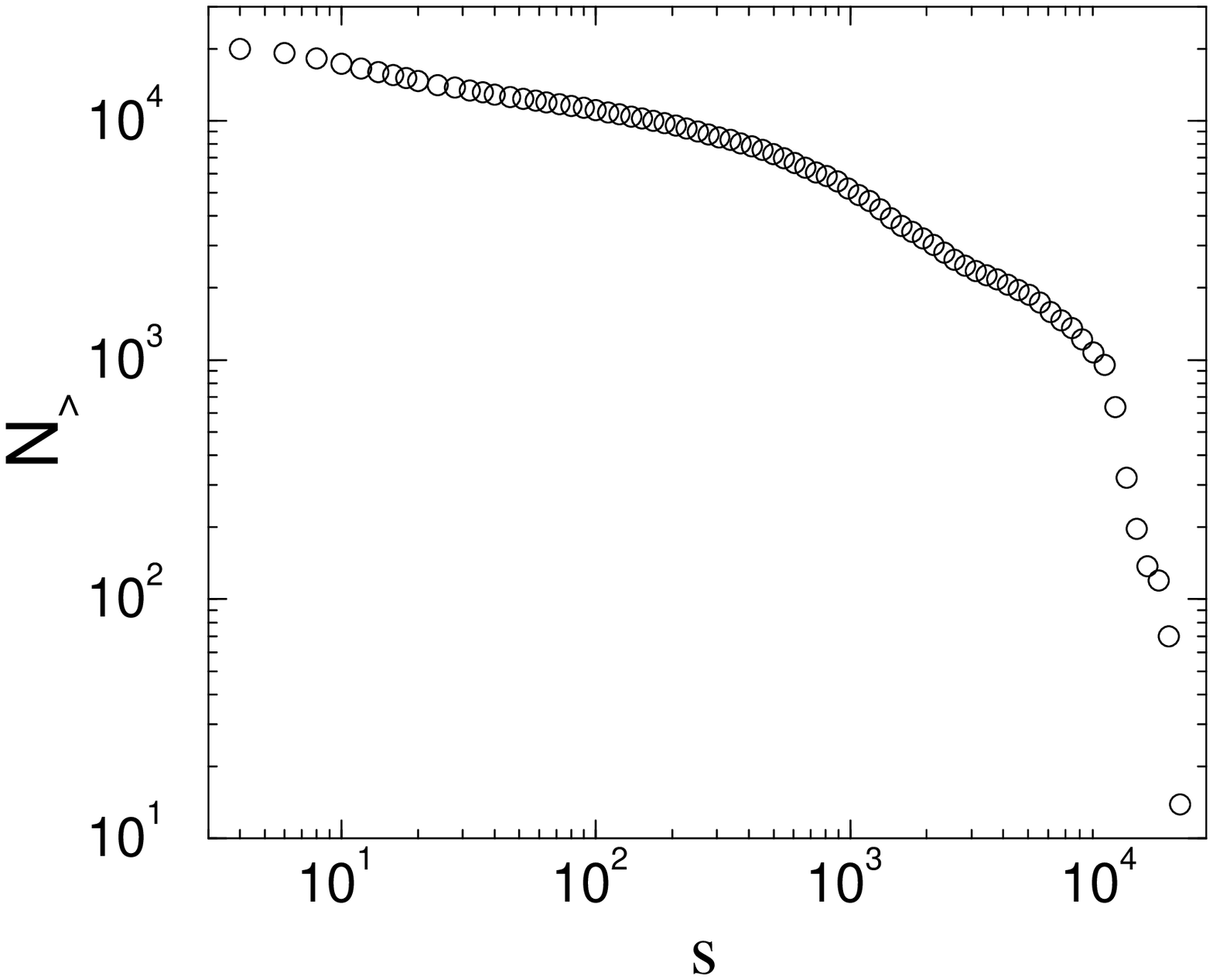}
\end{figure}
\end{minipage}

\begin{minipage}[t]{5cm}
\begin{figure}
\epsfxsize=5cm \epsfbox{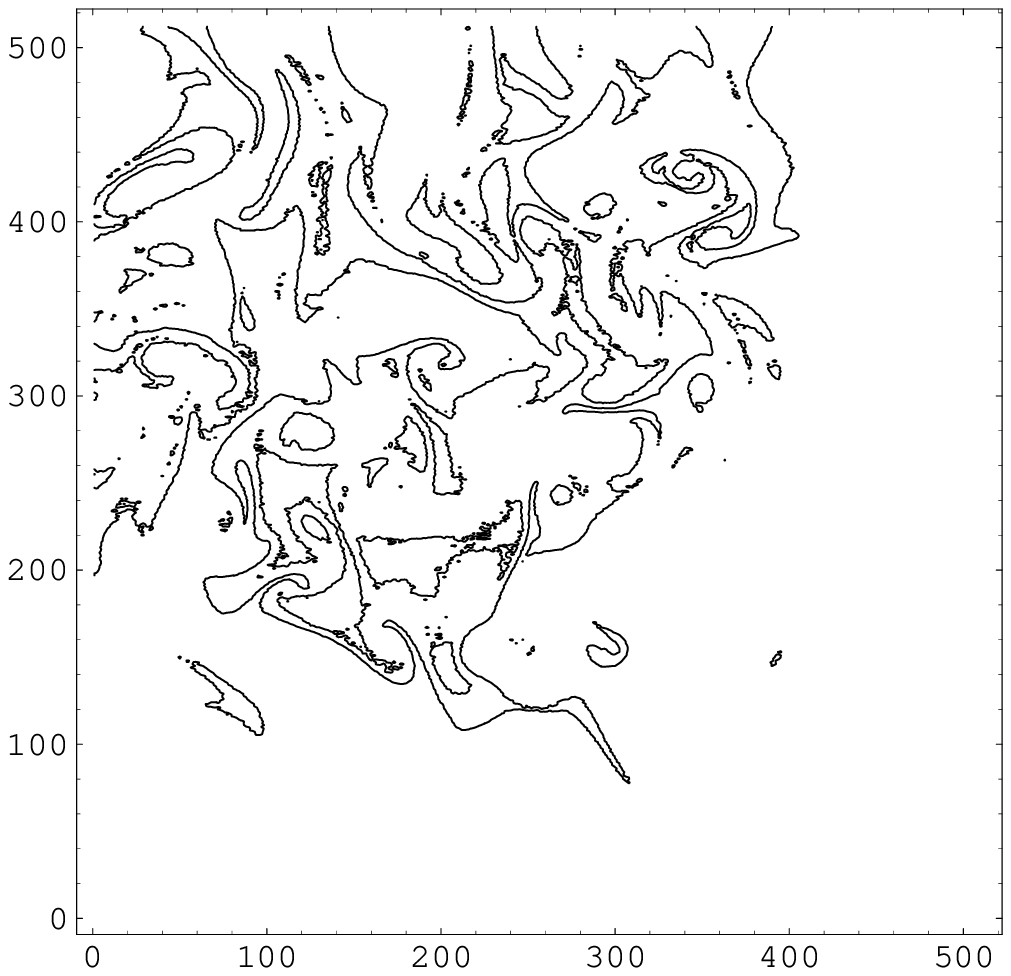}
\end{figure}
\end{minipage}
\hfill
\begin{minipage}[t]{6cm}
\begin{figure}
\epsfxsize=6cm \epsfbox{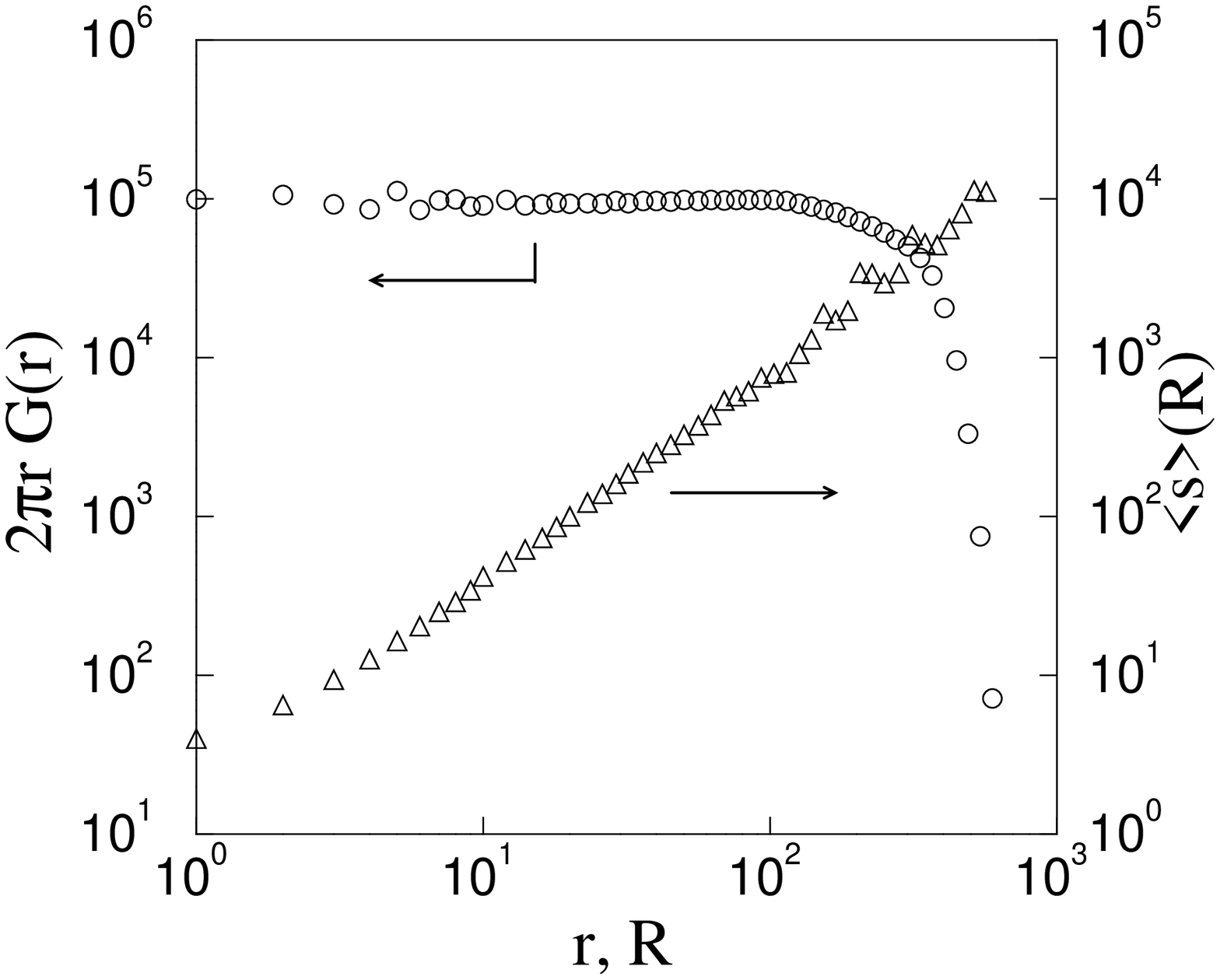}
\end{figure}
\end{minipage}
\hfill
\begin{minipage}[t]{6cm}
\begin{figure}
\epsfxsize=6cm \epsfbox{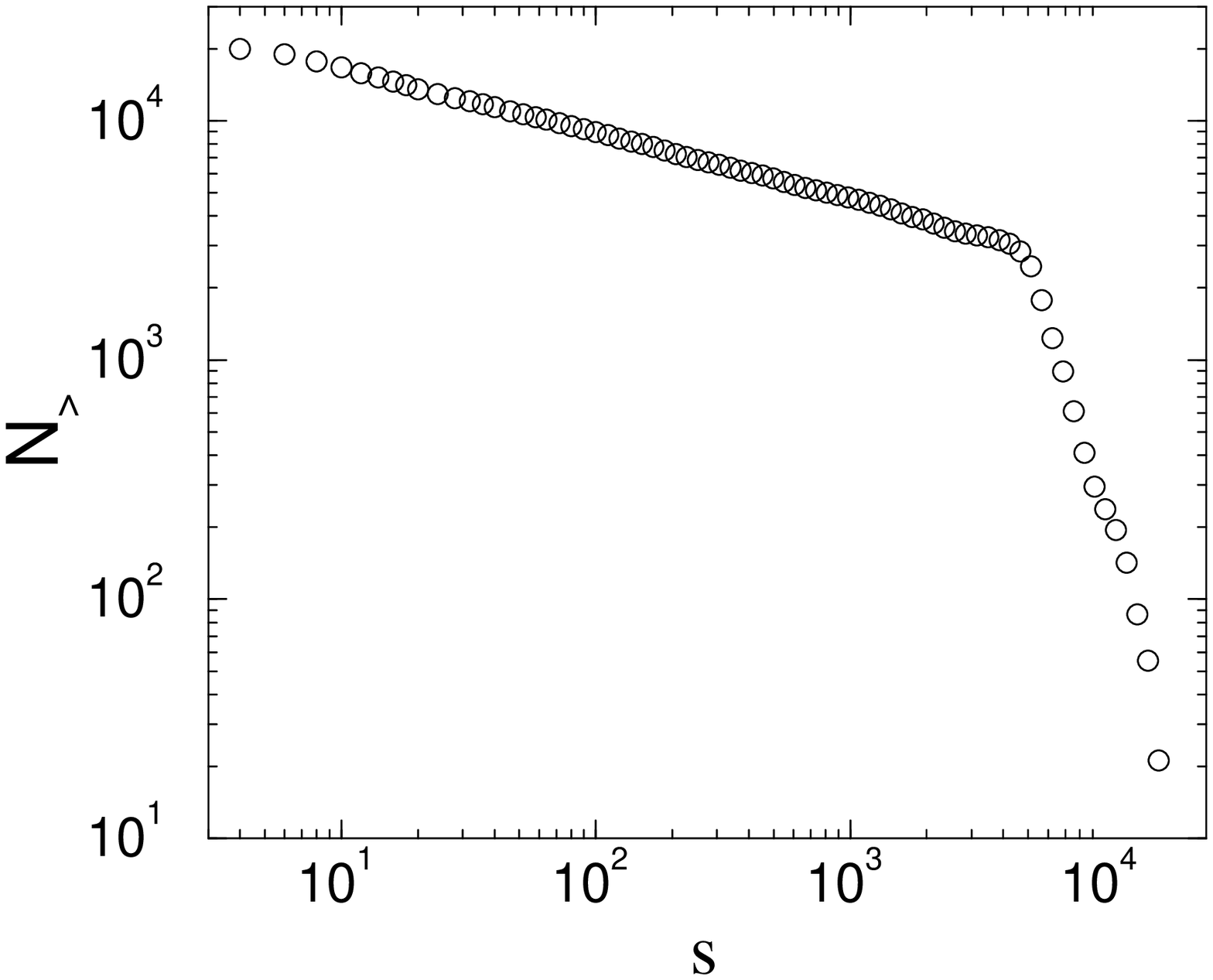}
\end{figure}
\end{minipage}

\noindent FIG.~1 The top row gives the level set and loop analysis of the 
scalar data in the direct cascade while the bottom row figures corresponds to 
scalar data obtained from the inverse cascade. All lengths ($r$,$R$, and $s$) 
are reported in units of one pixel; $G(r)$ is not normalized.

\begin{multicols}{2}

\noindent{\em i. Direct cascade -- } 
The loop correlation function shows the $1/r$ (i.e. $x_l=1/2$) behavior characteristic of 
rough fields for separations $r<100$ (see Fig.~1); in terms of the 
geometrical exponent $x_l$ we find from a linear least squares fit a value 
of $x_l=0.495(10)$. The rather large (compared to the inverse cascade data) crossover
region for $100<r<300$ which precedes the fall off  at $r>400$ (due to system size) 
might be indicative of an intermediate length scale at which the scaling of the 
loop ensemble changes. This is confirmed by the cumulative loop length distribution, $N_>(s)$.   

The graph of the mean loop length as a function of the radius,  Fig.~1, 
shows a scaling range for loop radii $3<R<300$. It leads to a fractal dimension 
for contour loops of $D = 1.20(3)$. The value of the exponent is estimated by 
comparing results from linear least-square fits to data within 
several different subintervals of the scaling range. 

The cumulative loop length distribution graph,  Fig.~1,  
shows scaling for $10<s<200$ after which there is a crossover to what 
looks like a power law with a smaller (more negative) exponent. The 
fall off  at lengths $s>7000$ occurs as the system size intervenes. 
The crossover occurs at length scales comparable to the effective dissipation 
scale which was identified from the scalar power spectrum in Ref.~\cite{tab_batch}.  
The exponent $\tau-2=0.19(2)$ is extracted from the scaling range, using the 
same procedure (sub-interval fits) as for $D_f$. 

Using the scaling relations \Eq{D-alpha}, we extract the value of the 
affine exponent $\alpha=0.60(6)$ from the measured $D$,  and 
$\alpha=0.53(8)$ from $\tau-2$. The fact that they match 
within error bars indicates that the loop data is consistent with 
{\em simple scaling} of the tracer field below the effective dissipation scale.
 
Above the dissipation scale Batchelor (logarithmic) scaling is expected to hold
\cite{tab_batch,batch}.  This 
implies $\alpha=0$ and is not well supported by the loop analysis. Namely, we 
would expect no crossover for  $x_l$ which is universal, and (from \Eq{D-alpha})
a crossover of $D$ to higher and $\tau-2$ to  a lower value.  
This is qualitatively consistent with the loop data (accept for $x_l$) 
but the actual numbers are off.  We
conclude that if the Batchelor regime is reached it occurs for a narrow range 
of scales which is not resolved by the loop analysis.  

Additional checks on the critical nature of the loop ensemble is provided by measuring 
the distribution of loop radii, $N_>(R)$. $N_>(R)$ is  the number of 
loops whose radius is larger than $R$ and  $N_>(R)\sim R^{-D(\tau-2)}$ follows 
from \Eq{int}. In the $N_>(R)$ plot we again see  a crossover occurring at 
$R\approx 60$, which from the $\left<s\right>(R)$ plot in Fig.~1 corresponds 
to $s\approx 300$, consistent with the $N_>(s)$ plot. For loops with $3<R<40$ 
we measure $D(\tau-2)=0.22(1)$ by fitting subintervals to a power law, 
while from the previous measurements of $D$ and $\tau-2$, $D(\tau-2)=0.23(3)$ 
follows.

\noindent {\em ii. Inverse cascade -- }
Loop analysis of the inverse energy cascade  images reveals a very 
different picture. The scaling is much better as no crossover scale 
is detected. The geometrical exponents can be 
extracted reliably from the data and they do {\em not} satisfy the scaling 
relations \Eq{D-alpha}.     

The loop correlation function decays with distance as a power law, 
for $r<100$ (see  Fig.~1); 
the measured value of the exponent $x_l=0.485(5)$ is consistent
with the universal value $x_l=1/2$.  
The $\left< s\right> (R)$ plot shows a scaling range for $10<R<100$, from which 
we extract $D=1.30(1)$ (see  Fig.~1). 
The  $N_>(s)$ plot,  Fig.~1, has a rather remarkable 
scaling range for loop lengths $10<s<3000$ from 
which the geometrical exponent $\tau-2=0.273(4)$ is obtained.  
Assuming the validity of the affine scaling relations \Eq{D-alpha} the exponent 
$\alpha=0.40(2)$ follows from the value of $D$, while $\alpha=0.25(2)$ 
is the value implied by the measured $\tau-2$. Clearly, these two values are
different, thus signaling the breakdown of simple scaling. We conclude that 
the scalar field is {\em intermittent}. 

To check that the loop ensemble is described by a joint distribution function 
that has a scaling form, \Eq{nsal}, we check for the validity of the scaling 
relation \Eq{scalerel1}. From the measured values of $D$ and $\tau-2$ we have
$D(\tau-3) = 0.95(1)$ while we expect $2-2x_l=1$ from the universal value 
$x_l=1/2$; $2-2x_l=1.03(1)$ is the value that follows from the $G(r)$ measure.
Furthermore we measure the combination $D(\tau-2)$ from $N_>(R)$.
We find $D(\tau-2)=0.344(4)$ (scaling range $3<R<100$), while the measures used 
above yield $D(\tau-2)=0.355(8)$. 
Taken together these results give strong support to the assumed critical 
nature of the contour loop ensemble of the scalar data in the inverse 
energy cascade regime.      

In conclusion we have shown that a scaling analysis of contour loops  of a 
passive scalar field advected by turbulent 
flow is a useful geometrical measure for studying its fluctuations. 
The loop analysis successfully distinguishes between different flow regimes and  
indicates the breakdown of simple scaling in the case of the 
inverse cascade. Even though the scalar field is intermittent in this case,  
the loop ensemble is critical and characterized by two 
independent geometrical exponents ($D$ and $\tau$). 
The direct cascade data reveals an intermediate length scale 
which can be identified with an effective dissipation length. 
Below this dissipation scale simple scaling of the scalar field is 
favored by the  loop analysis,  
while the data above this scale is not indicative of a Batchelor regime with 
logarithmic structure functions. 

The loop analysis described here has an advantage over 
structure-function measurements 
as the geometrical exponents are sensitive to higher moments 
of the concentration difference and they capture the
spatial structure of the fluctuations more fully by incorporating 
connectedness properties.  As three dimensional data becomes more 
available extensions of these ideas to contour {\em surfaces} 
will be a promising  direction for future research. 

We would like to thank P. Castiglione for valuable discussions regarding the 
experiments, and M.-C. Jullien and P. Tabeling for supplying the data 
sets. This work was supported by the NSF, award No.~DMR-9984471 (JK).

\end{multicols} 

\end{document}